\documentclass[11pt,final,onecolumn]{IEEEtran}

\def\editmode{0}

\def\bibfilenames{WISENET}

\if\editmode1  
\usepackage[backend=bibtex,style=alphabetic,sorting=debug]{biblatex}
\DeclareFieldFormat{labelalpha}{\thefield{entrykey}}
\DeclareFieldFormat{extraalpha}{}
\bibliography{\bibfilenames}
\newcommand{\cmt}[1]{\noindent\textcolor{lightgreen}{\underline{[#1]}}} 
\newcommand{\hc}[1]{\textcolor{blue}{#1}} 
\newenvironment{myitemize}{\begin{itemize}}{\end{itemize}}
\newcommand{\myitem}{\item}

\else
\usepackage{cite}
\newcommand{\cmt}[1]{} 
\newcommand{\hc}[1]{\textcolor{black}{#1}} 
\newenvironment{myitemize}{}{}
\newcommand{\myitem}{}

\fi 

\usepackage{fixltx2e}

\usepackage{graphicx}

\usepackage{subcaption}              

\usepackage[utf8]{inputenc}

\usepackage{amsfonts}
\usepackage{amsmath}
\usepackage{mathtools}

\usepackage{amssymb}

\usepackage{bm}

\usepackage{color,verbatim}
\usepackage{multirow}
\usepackage{accents}

\usepackage{theoremref}

\usepackage{url}

\newcounter{rulecounter}
\newcommand{\resetrule}{ \setcounter{rulecounter}{0}}
\resetrule

\newsavebox{\selvestebox}
\newenvironment{colbox}[1]
  {\newcommand\colboxcolor{#1}%
   \begin{lrbox}{\selvestebox}%
   \begin{minipage}{\dimexpr\columnwidth-2\fboxsep\relax}}
  {\end{minipage}\end{lrbox}%
   \begin{center}
   \colorbox{\colboxcolor}{\usebox{\selvestebox}}
   \end{center}}

\definecolor{orange}{rgb}{1,0.8,0}
\definecolor{gray}{rgb}{.9,0.9,0.9}
\definecolor{darkgray}{rgb}{.3,0.3,0.3}
\definecolor{darkblue}{rgb}{.1,0.0,0.3}
\definecolor{lightblue}{rgb}{0.7,0.7,1}
\definecolor{lightred}{rgb}{1,0.7,.7}
\definecolor{purple}{RGB}{204,153,255}
\definecolor{lightgray}{rgb}{.95,0.95,0.95}
\definecolor{lightgreen}{rgb}{0.3,0.5,0.3}
\definecolor{darkgreen}{rgb}{0.05,0.3,0.05}

\newcommand{\ra}{$\rightarrow$~}

\newcommand{\brackets}[1]{\left\{#1\right\}}

\newcommand{\tbm}[1]{{\tilde{\bm #1}}}
\newcommand{\cbm}[1]{{\check{\bm #1}}}
\newcommand{\hbm}[1]{{\hat{\bm #1}}}

\newcommand{\rfield}{\mathbb{R}}

\newcommand{\transpose}{^T}
 \newcommand{\define}{\triangleq}

\newcommand{\expected}[1]{\mathop{\textrm{E}}\brackets{#1} }

\newcommand{\minimize}{\mathop{\text{minimize}}}

\newcommand{\st}{\mathop{\text{s.t.}}}

\newtheorem{myproposition}{Proposition}
\newtheorem{myremark}{Remark}
\newtheorem{myproblemstatement}{Problem Statement}
\newtheorem{mylemma}{Lemma}
\newtheorem{mytheorem}{Theorem}
\newtheorem{mydefinition}{Definition}
\newtheorem{mycorollary}{Corollary}

\begin{document}
\renewcommand{\define}{{:=}}
\renewcommand{\transpose}{^\top}

\newcommand{\sourcenum}{{\hc{ S}}}
\newcommand{\sourceind}{{\hc{ s}}}
\newcommand{\transmitpsd}{{\hc{\Upsilon}}}
\newcommand{\receivedpsd}{{\hc{\Psi}}}
\newcommand{\measpsdmat}{{\tbm \truepsd} }
\newcommand{\augmeaspsdmat}{{\cbm \truepsd} }
\newcommand{\gridpoint}{{\hc{\bm \xi}}}
\newcommand{\nnfun}{\hc{p_{\bm w}}}

\newcommand{\truepsd}{\hc{\Psi}}
\newcommand{\mask}{\hc{M}}
\newcommand{\samplingfactor}{{\hc{ \gamma}}}

\newcommand{\pars}{\hc{\bm w}}
\newcommand{\encfun}{\hc{\epsilon}_{\pars}}
\newcommand{\decfun}{\hc{\delta}_{\pars}}
\newcommand{\code}{\hc{\bm \lambda}}
\newcommand{\latentnum}{\hc{N_\lambda}}

\newcommand{\auxinmat}{{\hc{\bm\Phi}}}
\newcommand{\auxlayerin}{\auxinmat^{(I)}}
\newcommand{\auxlayerout}{\auxinmat^{(O)}}
\newcommand{\inchind}{{\hc{c_\text{in}}}}
\newcommand{\inchnum}{{\hc{C_\text{in}}}}
\newcommand{\outchind}{{\hc{c_\text{out}}}}
\newcommand{\outchnum}{{\hc{C_\text{out}}}}

\newcommand{\layerfun}{\hc{p}}
\newcommand{\layerind}{{\hc{l}}}
\newcommand{\layernum}{\hc{L}}
\newcommand{\layerparnot}[2]{^{\hc{(}#1\hc{)}}_{#2}}
\newcommand{\psdvarmat}{\hc{\bm \chi}}

\title{Data-Driven Spectrum Cartography via \\Deep Completion Autoencoders}
\author{Yves Teganya$^{1,2}$ and Daniel Romero$^1$\\
  $^1$ Dept. of Information and Communication Technology, University
  of Agder, Norway.\\
  $^2$ Intelligent Signal Processing and Wireless Networks Laboratory (WISENET)
  \thanks{
Emails: \{yves.teganya, daniel.romero\}@uia.no. 
This work was supported by the Research Council of Norway through the FRIPRO TOPPFORSK grant 250910/F20 and
the INDNOR program under the LUCAT project. The authors thank
Prof. Baltasar Beferull-Lozano for funding and administrative support.
 }
}


\markboth{TECHNICAL REPORT, \today}{One, Two, and
Three: IEEE Journal Paper Template}

\maketitle
\begin{abstract}
Spectrum maps, which provide RF spectrum metrics such as power
spectral density for every location in a geographic area, find
numerous applications in wireless communications such as interference
control, spectrum management, resource allocation, and network
planning to name a few. Spectrum cartography techniques construct
these maps from a collection of measurements collected by spatially
distributed sensors. Due to the nature of the propagation of
electromagnetic waves, spectrum maps are complicated functions of the
spatial coordinates. For this reason, model-free approaches have been
preferred. However, all existing schemes rely on some interpolation
algorithm unable to learn from data. This work proposes a novel
approach to spectrum cartography where propagation phenomena are
learned from data. The resulting algorithms can therefore construct a
spectrum map from a significantly smaller number of measurements than
existing schemes since the spatial structure of shadowing and other
phenomena is previously learned from maps in other
environments. Besides the aforementioned new paradigm, this is also
the first work to perform spectrum cartography with deep neural
networks. To exploit the manifold structure of spectrum maps, a deep
network architecture is proposed based on completion
autoencoders. 

\end{abstract}
\section{Introduction}
\label{sec:introduction}
\cmt{Motivation}
\begin{myitemize}
\myitem\cmt{cartography overview}Spectrum cartography constructs maps
of RF channel metrics such as received signal power, interference
power, power spectral density (PSD), electromagnetic absorption, or channel
gain; see e.g.~\cite{alayafeki2008cartography,bazerque2010sparsity,jayawickrama2013compressive}.
\myitem\cmt{motivating applications}%
\begin{myitemize}%
\myitem\cmt{source localization}Besides applications like source
localization~\cite{bazerque2010sparsity} or radio
tomography~\cite{patwari2008correlated,romero2018blind},
\myitem\cmt{communications}spectral maps find a myriad of applications
in wireless communications such as network planning, interference
coordination, power control, spectrum management, resource allocation,
handoff procedure design, dynamic
spectrum access, and cognitive
ratio~\cite{grimoud2010rem,dallanese2011powercontrol,romero2017spectrummaps}.
\end{myitemize}%
\myitem\cmt{procedure}Spectrum maps are constructed from measurements
acquired by spectrum sensors or mobile devices.
\end{myitemize}

\cmt{Literature review}
\begin{myitemize}
\myitem\cmt{power maps}Most approaches are based on some interpolation
algorithm. For example, 
\begin{myitemize}
\myitem\cmt{kriging}power maps have been constructed through
kriging~\cite{alayafeki2008cartography,boccolini2012wireless},
dictionary learning~\cite{kim2011link,kim2013dictionary},
\myitem\cmt{compress. sensing, dict. lear., matrix
  completion}compressive sensing~\cite{jayawickrama2013compressive},
\myitem\cmt{bayes. models, radial basis functions, and kernel
  methods}Bayesian models~\cite{huang2015cooperative}, matrix
completion~\cite{ding2016cellular}, and kernel
methods~\cite{hamid2017non,teganya2019locationfree}.
\end{myitemize}%
\myitem\cmt{PSD maps}%
\begin{myitemize}%
\myitem\cmt{sparsity}PSD maps have also been
constructed by exploiting the sparsity of power across space and
frequency~\cite{bazerque2010sparsity} as well as by applying
 \myitem\cmt{kernel methods}thin-plate spline
regression~\cite{bazerque2011splines} and kernel-based
learning~\cite{romero2017spectrummaps, bazerque2013basispursuit}.
\end{myitemize}
\myitem\cmt{other metrics \ra channel gain maps}Metrics other than
power and PSD have also been mapped in the literature. For
example,~\cite{kim2011cooperative,romero2018blind,lee2018adaptive}
are capable of constructing channel gain maps. 
\myitem\cmt{Limitations}%
\begin{myitemize}%
\myitem\cmt{learn from experience}Unfortunately, none of the existing
approaches can learn from data. This means that they fail to learn the
characteristics of the propagation phenomena and, therefore, a
substantial performance improvement is expected if such knowledge can
be incorporated.
\end{myitemize}
\end{myitemize}

\cmt{Contributions}
\begin{myitemize}
\myitem\cmt{paradigm}To address this limitation, the first
contribution of this work is a data-driven paradigm for spectrum
cartography.  Specifically, it proposes learning the spatial
features of the relevant propagation phenomena such as shadowing,
reflection, and diffraction using a data set of past
measurements. Intuitively, leveraging these learned features can
significantly reduce the number of measurements required to attain a
target performance. This aspect is critical since all measurements
need to be collected in a sufficiently short time since the mapped
metric is subject to temporal variations in real-world scenarios.
\myitem\cmt{DNN algorithm}The second contribution comprises a spectrum
cartography algorithm to construct PSD maps  relying on a deep
neural network. Although several approaches for applying this class of
networks are discussed, the most natural one relies on a spatial
discretization of the area of interest.  The resulting tensor
completion task is addressed by means of a \emph{completion network}
architecture with an encoder-decoder structure that capitalizes on the observation that spectrum maps lie close to a low-dimensional
manifold embedded in a high-dimensional space. Our experiments reveal
that the performance of such algorithm beats the  state-of-the-art
alternatives.  \myitem\cmt{code and data}Finally, all code,
 trained networks, and the data set constructed for this work will
be posted at the authors' web sites.
\end{myitemize}

\cmt{emphasize novelty}The novelty of this work is twofold.
    \begin{myitemize}%
      \myitem\cmt{data-driven}First, this is the first work to propose
      a data-driven spectrum cartography approach.  \myitem\cmt{deep
        learning}Second, this is the first work to propose a deep
      learning algorithm for spectrum cartography.
    \end{myitemize}

\cmt{Paper structure}The rest of this report is organized as follows.
Sec.~\ref{sec:model} describes the problem of PSD
cartography. Sec.~\ref{sec:propmethod} presents the aforementioned
data-driven spectrum cartography paradigm and proposes a deep neural
network architecture based on completion autoencoders. Simulations and
conclusions are respectively provided in Secs.~\ref{sec:numtest}
and~\ref{sec:conclusion}.

\cmt{notation}\emph{Notation:}
    \begin{myitemize}
      \myitem\cmt{cardinality}$|\mathcal{A}|$ denotes the cardinality
      of set $\mathcal{A}$.  \myitem\cmt{i,j-th entry of
        matrix/tensor}$[\bm A]_{i,j}$ is the $(i,j)$-th entry of
      matrix $\bm A$, whereas $[\bm B]_{i,j,k}$ is the $(i,j,k)$-th
      entry of tensor $\bm B$.  \myitem\cmt{transpose}Finally, $\bm
      A^\top$ is the transpose of matrix $\bm A$.
    \end{myitemize}
\section{Model and Problem Formulation}
\label{sec:model}
\cmt{signal propagation overview}
\begin{myitemize}
  \myitem\cmt{sources}Consider $\sourcenum$ sources located in a
  geographical region of interest $\mathcal{X}\subset\rfield^2$ and
  operating on a certain frequency band. Let
  $\transmitpsd_{\sourceind}(f)$ denote the transmit PSD of the $\sourceind$-th source and
  \myitem\cmt{channel}let $H_{\sourceind}(\bm x, f)$ represent the
  frequency response of the channel between the $\sourceind$-th source
  and a sensor with an isotropic antenna located at $\bm x\in
  \mathcal{X}$. For simplicity, assume that 
  small-scale fading has been averaged out; see also
  Remark~\ref{rem:fading}.  \myitem\cmt{temporally stationary}Both
  $\transmitpsd_{\sourceind}(f)$ and $H_{\sourceind}(\bm x, f)$ are
  assumed to remain constant over time, a realistic assumption
  provided that the measurements described below are collected in an
  interval of shorter length than the channel coherence time and 
  time scale of changes in   $\transmitpsd_{\sourceind}(f)$. 

\myitem\cmt{measurement model}
      \begin{myitemize}
        \myitem\cmt{test location}If the $\sourcenum$ signals are
        uncorrelated, the PSD at $\bm x \in \mathcal{X}$ is 
$     \receivedpsd(\bm x, f)=\textstyle\sum_{\sourceind=1}^{\sourcenum} \transmitpsd_{\sourceind}(f)\vert H_{\sourceind}(\bm x, f) \vert^2 + \upsilon(\bm x, f)$
    with $\upsilon(\bm x, f)$ the noise PSD of a generic sensor at
    location $\bm x$, which models thermal noise, background radiation
    noise, and interference from remote sources.
    \myitem\cmt{measurement locations}A certain number of devices,
    such as mobile users in a cellular communication network or
    spectrum sensors, collect PSD measurements $\{ \tilde \receivedpsd
    (\bm x_n, f)\}_{n=1}^N$ at $N$ locations $\{\bm x_n\}_{n=1}^N
    \subset \mathcal{X}$ and finite set of frequencies $f\in
    \mathcal{F}$; see also Remark~\ref{rem:locations}.  These
    measurements can be obtained using e.g. periodograms or spectral
    analysis methods
    such as the Bartlett or Welch method~\cite{stoica2005}.
      \end{myitemize}
      
      \myitem\cmt{fusion center}These measurements are sent to a
      fusion center, which may be e.g. a base station, a mobile user, or a
      cloud server, depending on the application. 
    \end{myitemize}%
\cmt{spectrum cartography problem}%
\begin{myitemize}%
      \myitem\cmt{given}Given $\{(\bm x_n, \tilde \receivedpsd(\bm
      x_n, f) ),~n=1,\ldots,N,~f\in\mathcal{F}\}$,
      \myitem\cmt{requested}the fusion center must obtain an estimate
      $\hat{\receivedpsd}(\bm x, f)$ of $\receivedpsd(\bm x, f)$ at
      every location $\bm x \in \mathcal{X}$ and frequency $f\in
      \mathcal{F}$.  \myitem\cmt{terminology}In spectrum cartography,
      function $\receivedpsd(\bm x, f)$ is typically referred to as
      the \emph{true map}, whereas $\hat{\receivedpsd}(\bm x, f)$ is
      the \emph{estimated map} or \emph{map estimate}. The algorithm
      or rule that provides a map estimate, which in this work is a
      neural network, is termed \emph{map estimator}.
          \end{myitemize}%
    \cmt{challenge}%
    \begin{myitemize}%
      \myitem\cmt{minimize error}The challenge is to exploit the spatial
      structure of propagation phenomena so that the estimation
      error, quantified e.g. as $\sum_f\int_\mathcal{X} |
      \receivedpsd(\bm x, f) - \hat\receivedpsd(\bm x, f) |^2d\bm x$,
      is minimized for a certain $N$ or, alternatively, the minimum
      $N$ required to guarantee a target estimation error is
      minimized.
      
      \myitem\cmt{existing approaches}To the best of our knowledge,
      all existing approaches to spectrum cartography are based on
      interpolation algorithms that do not learn from data. 
      \myitem\cmt{outlook}In contrast, the next section develops
      a novel data-driven methodology that learns the aforementioned
      structure from a data set. 
      \begin{myremark}
        \label{rem:fading}Sensors must determine their locations
        $\{\bm x_n\}_n$ with an error sufficiently small relative to
        the scale of spatial variability of $\truepsd(\bm x,f)$ across
        $\mathcal{X}$. Thus, estimating  small-scale fading is 
        more challenging than estimating shadowing since the coherence
        distance of the former is comparable to the wavelength and
        typical communication bands of interest have wavelengths in
        the order of centimeters.
      \end{myremark}
      \begin{myremark}
        \label{rem:locations}The number of measurement locations may be
        significantly larger than the number of sensors if the sensors
        move. Measurements collected at different locations may be
        useful to estimate a spectrum map so long as the difference
        between measurement instants is small relative to the time
        scale of the variations of the PSD map. The latter is highly
        dependent on the specific application. For example, one
        expects significant variations in DVB-T bands to occur in the
        scale of several months, whereas PSD maps in LTE bands may
        change in the scale of milliseconds due to power control,
        mobility, and interference.
      \end{myremark}
      
    \end{myitemize}

    \section{Proposed Data-Driven Cartography}
    \label{sec:propmethod}
    \cmt{section overview}This section introduces a data-driven
    paradigm for spectrum cartography and develops a deep learning
    algorithm that abides by this principle. To this end,
    Sec.~\ref{sec:tensorcompletion} starts by reformulating the
    problem at hand as a tensor-completion task amenable to
    application of deep neural networks. Subsequently,
    Sec.~\ref{sec:missing} addresses unique aspects of tensor/matrix
    completion via deep learning. Finally, Secs.~\ref{sec:realworld}
    and~\ref{sec:convautoenc} respectively describe how a deep neural
    network can be trained to learn the spatial structure of
    propagation phenomena and how this task can be addressed via the
    notion of \emph{completion autoencoders}.
    
    \subsection{Spectrum Cartography as a Tensor Completion Task}
    \label{sec:tensorcompletion}
    \cmt{motivation}
    \begin{myitemize}
      \myitem\cmt{challenge}Observe that the value of $N$ depends on the
    number and movement of the sensors relative to  the time-scale of
    temporal  changes in $\receivedpsd(\bm x,f)$;
    cf. Sec.~\ref{sec:model}.
    \myitem\cmt{Possible approaches}%
        \begin{myitemize}\myitem\cmt{many vs one}          
        \begin{myitemize}\myitem\cmt{separate estimators}In principle, one could think of using a
    separate map estimator for each possible value of $N$. Each
    estimator could be relatively simple since it would always take the
    same number of inputs. However, such an approach would be highly
    inefficient in terms of memory, computation, and prone to erratic
    behavior since each estimator is trained with a different data
    set.
    \myitem\cmt{one estimator}Thus, it is more practical to rely on a single estimator that
    can accommodate arbitrary values of $N$.    
        \end{myitemize}
\end{myitemize}%
        \begin{myitemize}%
          \myitem\cmt{deep learning}%
          \begin{myitemize}%
      \myitem\cmt{recurrent}A customary approach in deep learning
      for coping with inputs of variable lengths is through recurrent
      neural networks~\cite{mandic2001recurrent},\cite[Ch. 10]{goodfellow2016deep}. Unfortunately, besides the
      difficulties of training these networks, it is unclear how such
      an approach could effectively exploit spatial information.
      \myitem\cmt{tensor completion}For this reason, the selected
      approach in this work is to reformulate the cartography problem
      as a tensor completion task amenable to a solution based on a
      feedforward architecture~\cite[Ch. 6]{goodfellow2016deep}.
    \end{myitemize}
        \end{myitemize}
        \end{myitemize}

    \begin{figure}[t!]
    \centering
    \includegraphics[scale=0.3]{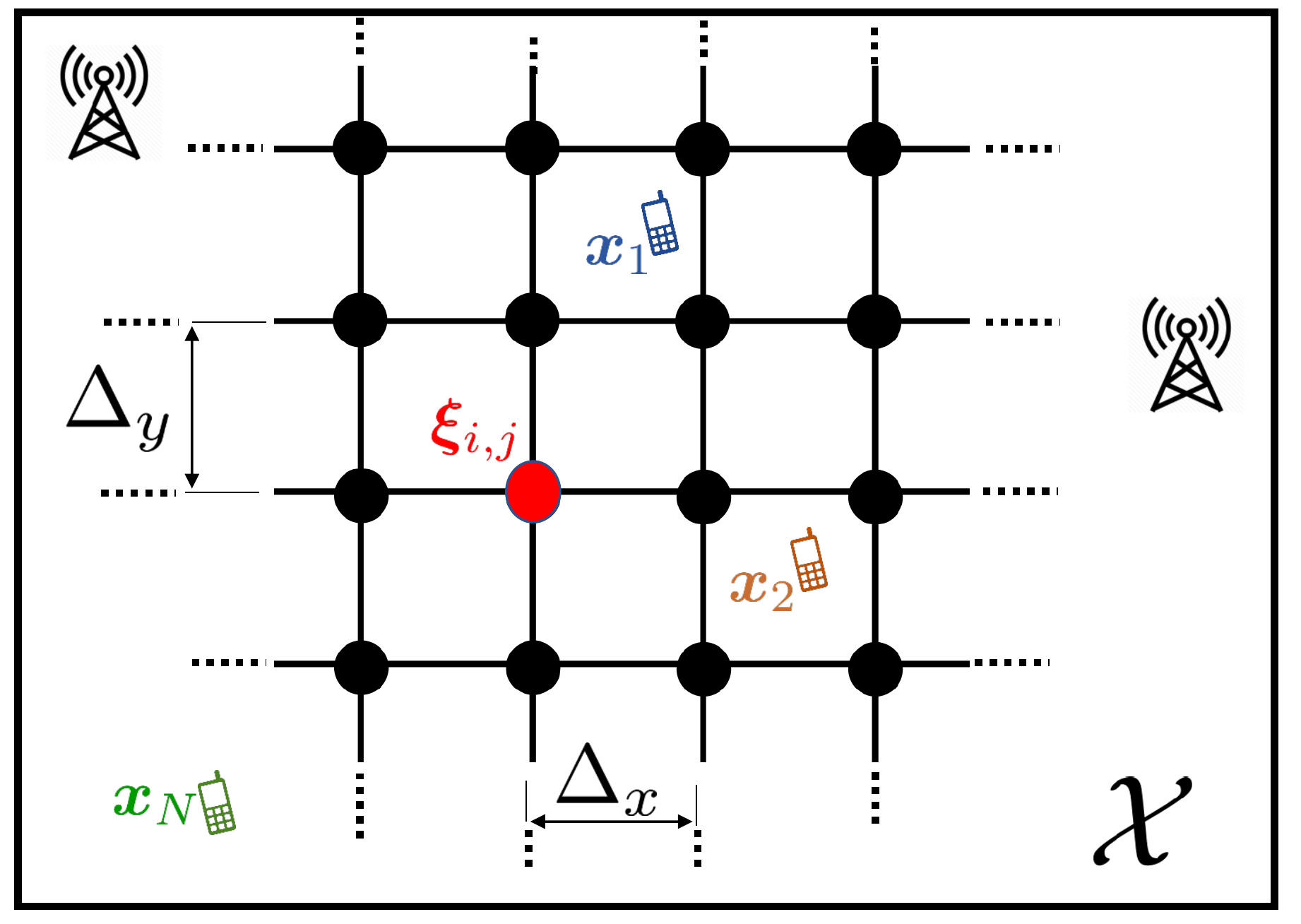}
    \caption{Model setup and area discretization.} 
    \label{f:setupwithgrid}
    \end{figure}

    \cmt{spatial discretization}To this end, one must discretize
    $\mathcal{X}$, a trick already applied in radio tomographic
    imaging~\cite{romero2018channelgain,hamilton2014modeling} and
    spectrum cartography~\cite{ding2016cellular}. To introduce the
    appropriate notation, it will be briefly outlined next.
      \begin{myitemize}%
          \myitem\cmt{2D grid}Define an $N_y \times N_x$ rectangular
          grid over $\mathcal{X}$, as depicted in
          Fig.~\ref{f:setupwithgrid}. This grid comprises points
          $\gridpoint_{i,j}$ evenly spaced by $\Delta_x$ and
          $\Delta_y$ along the $x$- and $y$-axes respectively, that
          is, the $(i,j)$-th grid point is given by
          $\gridpoint_{i,j}:=\left [i\Delta_x,~j\Delta_y\right]^\top$,
          with $~i=1,\ldots,N_y ,~j=1,\ldots,N_x$.  \myitem\cmt{assign
            meas. loc to grid pts}For future usage, define
          $\mathcal{A}_{i,j} \subset \{1,\ldots,N\}$ as the set
          containing the indices of the measurement locations assigned
          to the $(i,j)$-th grid point by the criterion of minimum
          distance, i.e., $n\in \mathcal{A}_{i,j}$ iff
          $||\gridpoint_{i,j}-\bm x_n||\leq ||\gridpoint_{i',j'}-\bm
          x_n||\forall i',j'$ with $i'\neq i$ and $j'\neq j$.
    \end{myitemize}

      \cmt{True map}This grid induces a discretization of $ \truepsd
      (\bm x, f) $ along the $\bm x$ variable. One can therefore
      collect the true PSD values at the grid points in matrix $\bm
      \truepsd (f) \in \rfield^{N_y \times N_x}$, $f\in \mathcal{F}$,
      whose $(i,j)$-th entry is given by $\left[ \bm \truepsd (f)
        \right]_{i,j}= \receivedpsd(\gridpoint_{i,j}, f)$. By letting
      $\mathcal{F}=\{f_1,\ldots,f_{N_f}\}$, it is also possible to
      stack these matrices along the third dimension to form the
      tensor $\bm \truepsd\in \rfield^{ N_y\times N_x\times N_f}$,
      where $\left[ \bm \truepsd \right]_{i,j,n_f}=
      \receivedpsd(\gridpoint_{i,j}, f_{n_f})$,
      $n_f=1,\ldots,N_f$. For short, the term \emph{true map} will
      either refer to $ \receivedpsd(\bm x, f)$ or $\bm \truepsd$.

      \cmt{measurements}Similarly, one can collect the measurements in
      a tensor of the same dimensions. 
    \begin{myitemize}
          \myitem\cmt{motivate aggregation}Informally, if the grid is
          sufficiently fine ($\Delta_x$ and $\Delta_y$ are
          sufficiently small), it holds that $\bm x_n \approx
          \gridpoint_{i,j}$ $\forall n \in \mathcal{A}_{i,j}$ and,
          correspondingly, $ \receivedpsd(\bm x_n, f) \approx
          \receivedpsd(\gridpoint_{i,j}, f)$ $\forall n \in
          \mathcal{A}_{i,j}$. It follows that,
$
       \receivedpsd(\gridpoint_{i,j}, f)\approx({1}/{\vert \mathcal{A}_{i,j} \vert)}\sum_{n \in \mathcal{A}_{i,j} }\receivedpsd(\bm x_n, f)
       $
          whenever $\vert \mathcal{A}_{i,j} \vert\geq 1 $.       
          \myitem\cmt{aggregation}Therefore, it makes sense to aggregate all the measurements
      assigned to $\gridpoint_{i,j}$ as\footnote{For simplicity, the
        notation implicitly assumes that $\bm x_n\neq
        \gridpoint_{i,j}$ $\forall n,i,j$, but this is not a requirement.}
$
       \tilde \receivedpsd(\gridpoint_{i,j}, f)\define ({1}/{\vert
         \mathcal{A}_{i,j} \vert})\sum_{n \in \mathcal{A}_{i,j} }
       \tilde \receivedpsd(\bm x_n, f). 
$
       \myitem\cmt{misses} Conversely, when $|\mathcal{A}_{i,j}|=0$,
       there are no measurements associated with
       $\gridpoint_{i,j}$, in which case one says that there is a
       \emph{miss} at        $\gridpoint_{i,j}$.
       \myitem\cmt{sampled map matrix}Upon letting $\Omega \subset \{1,\ldots,N_y\}\times \{1,\ldots, N_x\}$ be 
such that $(i,j) \in \Omega$ iff $\vert \mathcal{A}_{i,j} \vert > 0$, all (possibly aggregated)
       measurements $\tilde \receivedpsd(\gridpoint_{i,j}, f)$ can be
       collected in  $\tilde {\bm \truepsd} (f) \in \rfield^{N_y
  \times N_x}$, defined as $[ \tilde {\bm \truepsd} (f)
         ]_{i,j}=                   \tilde
       \receivedpsd(\gridpoint_{i,j}, f)$ if $
                  (i,j) \in \Omega$ and $[ \tilde {\bm \truepsd}
         (f) ]_{i,j}=0$ otherwise. 
\myitem\cmt{filling}Note that misses  have been
filled with zeroes, but other values can be used. 

    
      \myitem\cmt{error sources}When $(i,j)\in \Omega$,
      the values of $ [ \tilde {\bm \truepsd} (f) ]_{i,j}$
      and $ [ {\bm \truepsd} (f) ]_{i,j}$ differ due to the
      error introduced by the spatial discretization as well as due to
      the measurement error incurred when measuring
      $\truepsd(\bm x_n,f),~n\in \mathcal{A}_{i,j}$. The latter is
      caused mainly by the finite time devoted by sensors to take 
      measurements.

      \myitem\cmt{tensor}As before, the matrices
      $\tbm\truepsd(f),~f=1,\ldots,N_f$ can be stacked along the 3rd
      dimension to form  $\tbm \truepsd \in \rfield^{ N_y\times
      N_x\times N_f}$, where  $[\tbm \truepsd ]_{i,j,n_f} =
    [\tbm \truepsd (f_{n_f}) ]_{i,j}$. For short, this tensor will be
    referred to as the \emph{sampled map}. 
    \end{myitemize}


    \cmt{Problem reformulation}The cartography problem stated in
    Sec.~\ref{sec:model} can now be  reformulated as,
    \begin{myitemize}%
      \myitem\cmt{given}given $\Omega$ and $\tbm \truepsd$,
      \myitem\cmt{req.}estimate $\bm \truepsd$. 
\end{myitemize}



    \subsection{Feedforward Completion Networks}
    \label{sec:missing}
    
\cmt{overview}The previous section reformulated the spectrum
cartography problem as a tensor completion task. Since conventional
neural networks cannot directly accommodate input misses and
set-valued inputs like $\Omega$, this section
explores the possibilities and motivates the adopted approach.

\cmt{DL refresh}But before that, a quick refresh on deep learning is
in order.  A feedforward deep neural network is a function $\nnfun$
that can be
\begin{myitemize}%
  \myitem\cmt{layers}%
expressed as the composition
$
\nnfun(\auxinmat)= \layerfun\layerparnot{\layernum}{\bm w_{\layernum}}(\layerfun\layerparnot{\layernum-1}{\bm w_{\layernum-1}}(\ldots\layerfun\layerparnot{1}{\bm w_{1}}(\auxinmat)))
 $
of \emph{layer} functions $\layerfun\layerparnot{\layerind}{\bm
  w_\layerind}$, where $\auxinmat$ is the input. Although there is no
commonly agreed definition of layer function, it is typically formed
by concatenating simple scalar-valued functions termed \emph{neurons}
that implement a linear function followed by a simple non-linear
function known as activation~\cite{goodfellow2016deep}.
\myitem\cmt{neural}The term \emph{neuron} stems from the resemblance
between these functions and certain simple functional models for natural
neurons.  \myitem\cmt{deep}Similarly, there is no consensus on which
values of $\layernum$ qualify for $\nnfun$ to be considered a
\emph{deep} neural network.  \myitem\cmt{parameters}With vector  $\bm
w_\layerind$ containing the parameters of the $\layerind$-th layer, the
parameters of the entire network can be collected in $\bm w \define [\bm w_1\transpose, \ldots,\bm
  w_{\layernum}\transpose]\transpose \in \rfield^{N_w}$.  \myitem\cmt{training}These
parameters are \emph{learned} using a \emph{training set} in a process termed
\emph{training}.

    \end{myitemize}%

\begin{myitemize}%
  \myitem\cmt{Training in our application \ra outlook}The rest of this
  section as well as Sec.~\ref{sec:realworld} carefully delineate how
  a deep neural network can be trained to perform data-driven spectrum
  cartography. Occasional references to works in areas such as
  collaborative filtering and image inpainting will provide insight
  and motivate the design decisions. On the other hand,
  Sec.~\ref{sec:convautoenc} will address the design of $\nnfun$.

\myitem\cmt{training data}Although the training set construction is
detailed in Sec.~\ref{sec:realworld}, suppose by now that a
set of $T$ \emph{training examples} $\{( \tbm \truepsd_t,
\Omega_t)\}_{t=1}^{T}$ is given. Here, $\{\tbm \truepsd_t\}_t$ is a
collection of sampled maps acquired in different environments and
$\Omega_t$ the corresponding sampling set.
\end{myitemize}%

\cmt{Dealing with misses}%
\begin{myitemize}%
  \myitem \cmt{maximize w.r.t. missing entries}The desired estimator
  should obtain $ \bm \truepsd$ as a function of $\tbm \truepsd$ and
  $\Omega$. But regular neural networks cannot directly accommodate
  set-valued inputs and  missing entries. For this
  reason, \cite{fan2017deep} proposes filling the
  \begin{myitemize}%
    \myitem\cmt{description}missing entries in $\tbm \truepsd$ by solving
    \begin{myitemize}%
      \myitem\cmt{training phase}
          \begin{myitemize}%
      \myitem\cmt{Obj}
      \begin{align}  \label{eq:matricomp}
\underset{\{\psdvarmat_t\}_t, \bm w}{ \text{minimize}}\quad
&\frac{1}{T}\sum_{t=1}^T
\left \Vert \mathcal{P}_{\Omega_t}\left( \psdvarmat_t - p_{\bm w}(\psdvarmat_t
)\right)  \right \Vert_F^2, \\
\st\quad& [\psdvarmat_t]_{i,j,n_f}=[\tbm \truepsd_t]_{i,j,n_f}~\forall
n_f,\forall (i,j)\in \Omega_t,
\nonumber
\end{align}
where   $||\bm A||_F^2\define
\sum_{i,j,n_f}[\bm A]_{i,j,n_f}^2$ is the Frobenius norm of  tensor $\bm A$ and 
$\mathcal{P}_{\Omega}(\bm A)$ is defined as $\left[ \mathcal{P}_{\Omega}(\bm A) \right]_{i,j,n_f}=
    [\bm A]_{i,j,n_f}$ if $(i,j) \in \Omega$ and
    $\left[ \mathcal{P}_{\Omega}(\bm A) \right]_{i,j,n_f}= 0$ otherwise.
    The map estimate produced by this
method is directly the minimizer $\psdvarmat_t$ of \eqref{eq:matricomp}.
\myitem\cmt{complexity control}Observe that
if there exists a value of $\bm w$ for which $\nnfun$ becomes the
identity map, i.e. $\psdvarmat= p_{\bm
  w}(\psdvarmat ),~\forall \psdvarmat$, then the optimum of
\eqref{eq:matricomp} is attained regardless of the value of the
entries $[\psdvarmat_t]_{i,j,n_f},~(i,j)\notin \Omega_t$, which would
render this estimator useless. Thus, some form of \emph{capacity/complexity
  control} is necessary~\cite{cherkassky2007}. For
instance, one can
(i)\cmt{constraints} impose constraints on $\bm w$, 
(ii)\cmt{regularizer} add a regularization term to the objective
function, or 
(iii)\cmt{architecture} 
limit capacity through the design of the network architecture.
Approach (iii)  will be discussed
      further in  Sec.~\ref{sec:convautoenc}.
To simplify the exposition, expressions in this report will not
display constraints or regularizers, but it is understood that the
user may include them if necessary.
    \end{myitemize}%

      \myitem\cmt{testing phase}After $\bm w=\hbm w$ has been obtained
      by applying \eqref{eq:matricomp} with sufficiently large $T$,
      one can complete further tensors $\tbm \truepsd_t$ by
      setting $\bm w$ in \eqref{eq:matricomp} to this
      learned vector $\hbm w$ and optimize only with respect to $
      \{\psdvarmat_t\}_t$, which is computationally simpler.

    \end{myitemize}%

\myitem\cmt{limitations \ra complexity \ra prohibitive}
\begin{myitemize}%
  \myitem\cmt{no. opt. var}The number of optimization variables in \eqref{eq:matricomp} is
  $N_w + N_xN_yN_fT$, where $N_w$ is the  length of $\bm w$. This number  is
   prohibitive for high  $T$, as required for
  training deep neural networks.  \myitem\cmt{testing
    phase}Besides, even with the aforementioned simplified
  approach that only optimizes with respect to $ \{\psdvarmat_t\}_t$,
  a large number of forward and backward backpropagation passes~\cite[Ch. 6]{goodfellow2016deep} are required to
  estimate each map. Thus, this approach is not suitable for
  real-time implementation, as required in spectrum cartography
  applications. 
    \end{myitemize}%

  \end{myitemize}%
  \myitem\cmt{ not minimize, feed zeroes or another filling value. }To
  alleviate this limitation, a simple alternative would be to just
  feed $\tbm \truepsd$ to the neural network and train by solving
      \begin{myitemize}%
      \myitem\cmt{description}
  \begin{myitemize}%
    \myitem\cmt{training}
    \begin{align}  \label{eq:matricompw}
      \underset{\bm w}{ \text{minimize}}\quad
      &\textstyle\frac{1}{T}\sum_{t=1}^T
      \left \Vert \mathcal{P}_{\Omega_t}\left( \tbm \truepsd_t- p_{\bm w}(\tbm \truepsd_t
      )\right)  \right \Vert_F^2.
    \end{align}
    Although the missing entries were filled with zeros in
    Sec.~\ref{sec:tensorcompletion}, one can alternatively use other real numbers. 
    \myitem\cmt{testing}After \eqref{eq:matricompw} is solved, $\tbm
    \truepsd$ can be completed just by evaluating $p_{\bm w}(\tbm
    \truepsd)$, 
    \end{myitemize}%
  \end{myitemize}%
    \begin{myitemize}%
      \myitem\cmt{strengths}
      \begin{myitemize}%
        \myitem\cmt{testing phase}which requires a single forward pass. 
        \myitem\cmt{vars}Besides, solving \eqref{eq:matricompw}
        involves just $N_w$ optimization variables.
      \end{myitemize}%
    \myitem\cmt{limitation}However, because the completion step
    $p_{\bm w}(\tbm \truepsd)$ does not involve $\Omega$, poor
    performance is expected since  the network cannot distinguish missing entries
    from  measurements close to the filling value.
    \end{myitemize}%

\myitem\cmt{natural units}In the application at hand, one could
circumvent this limitation by expressing the entries of $\tbm
\truepsd$ in natural power units (e.g. Watt) and filling the misses
with a negative number such as -1. Unfortunately, the usage of
finite-precision arithmetic would introduce large errors in the map
estimates and is problematic in our experience. For this reason,
expressing $\tbm \truepsd$ in logarithmic units such as dBm is
preferable. However, the problem of distinguishing missing entries
persists since logarithmic units are not confined to be non-negative.
    
  \myitem\cmt{complement input with a mask \ra add a channel}A more practical approach is to
          complement the input of the network with a binary mask that
          indicates which entries are observed, as proposed in the
          image inpainting literature~\cite{iizuka2017consistent}. 
  \begin{myitemize}%
    \myitem\cmt{mask}
In this case, the binary mask $\bm \mask_{\Omega} \in \{0,1\}^{N_y
  \times N_x}$ associated with the sampling set $\Omega$ is given by
$[\bm \mask_{\Omega}]_{i,j}=1$  if $
              (i,j) \in \Omega$ and $[\bm \mask_{\Omega}]_{i,j}=  0$ 
               otherwise.

\myitem\cmt{mask as a channel}To simplify notation, let  $\augmeaspsdmat\in
\rfield^{N_y\times N_x \times N_f+1}$ denote a tensor obtained by concatenating $\tbm \truepsd$ and
$\bm \mask_{\Omega}$ along the third dimension.
\myitem\cmt{training}The neural network can therefore be trained as
\begin{align}  
      \underset{\bm w}{ \text{minimize}}\quad
      &\textstyle \frac{1}{T}\sum_{t=1}^T
      \left \Vert \mathcal{P}_{\Omega_t}\left( \tbm \truepsd_t- p_{\bm w}(\augmeaspsdmat_t
      )\right)  \right \Vert_F^2
    \end{align}
    \myitem\cmt{testing}and, afterwards, a tensor $\tbm \truepsd$ can
    be completed just by evaluating $p_{\bm w}(\augmeaspsdmat )$.
    \myitem\cmt{strengths}Then, this scheme is simple to train,
    inexpensive to test, and exploits information about the location
    of the misses. 
\end{myitemize}%
\end{myitemize}%

\subsection{Learning in Real-World Scenarios}
\label{sec:realworld}

\cmt{data-driven}A key novelty in this work is to obtain map
estimators by learning from data. This section describes how to
construct a suitable training set in the application at hand.

\cmt{Alleviating Ill-Posedness\ra frequency separation}
\begin{myitemize}%
  \myitem\cmt{issue}The first consideration pertains to
  ill-conditioning issues arising when the number of frequencies $N_f$
  in $\mathcal{F}$ is large, as will typically be the case.  Suppose
  that the first layer of $\nnfun$ is fully connected and has $N_N$
  neurons. Its total number of parameters becomes $(N_yN_xN_f+1)N_N$
  plus possibly additional parameters of the activation
  functions. Other layers will experience the same issue to different
  extents.  Since $T$ must be comparable to the number of unknowns to
  train the network effectively, the impact of a large $N_f$ is to
  drastically limit the number of layers or neurons that can be used.
  \myitem\cmt{separation}
  \begin{myitemize}%
    
    \myitem\cmt{literature}Previous approaches in spectrum cartography
    experienced similar issues, which were often addressed by the
    introduction of parametric models along the frequency domain; see
    e.g.~\cite{bazerque2011splines,romero2017spectrummaps}.
    \myitem\cmt{idea}Although such an approach can be similarly
    adopted in the present work, thereby reducing the number of
    \emph{channels} at the neural network input from $N_f+1$ to a
    much smaller number, it will be argued next that directly separating the
    problem across frequencies may be preferable when training a deep
    neural network. The idea is that propagation phenomena at similar
    frequencies are expected to be similar. Building upon this
    principle, $\nnfun$ can operate separately at each frequency
    $f$. This means that training can be accomplished through
    \myitem\cmt{train}
\begin{align} \label{eq:fsep}
  \underset{\bm w}{ \text{min.}}~
  &\frac{1}{T N_f}\sum_{t=1}^T\sum_{f\in \mathcal{F}}
  \left \Vert \mathcal{P}_{\Omega_t}\left( \tbm \truepsd_t(f)- p_{\bm w}(\augmeaspsdmat_t(f)
  )\right)  \right \Vert_F^2, 
\end{align}
    where $\augmeaspsdmat_t(f)\in \rfield^{N_y\times N_x \times 2}$ is
    a tensor with first frontal slab given by $\tbm \truepsd_t(f)$ and
    second frontal slab given by $\bm \mask_{\Omega_t}$.

    \myitem\cmt{strengths}
    \begin{myitemize}%
      \myitem\cmt{variables}Observe that the number of variables is
      now reduced by a factor of $N_f$ whereas \myitem\cmt{data
        examples} the ``effective'' number of training examples has been
      multiplied by $N_f$; cf. number of summands in
      \eqref{eq:fsep}. This is a drastic improvement especially
      when $N_f$ takes values such as 512 or 1024, as customary in
      spectral analysis.  Thus, such a frequency separation allows an
       increase in the number of neurons per layer or (typically more
      useful~\cite[Ch. 5]{goodfellow2016deep}) the total number of layers for a
      given $T$. Although such a network
      would not exploit structure across the frequency domain, the
      fact that it would be better trained  is likely
      to counteract this limitation in many setups. 
    \end{myitemize}%

          \end{myitemize}%

      \end{myitemize}%

\cmt{training approaches}The next step is to construct the data set,
for which three approaches are discussed next:
\subsubsection{Synthetic Training Data}
\begin{myitemize}%
  \myitem\cmt{idea}Since collecting a large number of training maps
  may be slow or expensive, one can instead generate maps using a
  mathematical model or simulator that captures the structure of the
  propagation phenomena; see
  e.g.~\cite{jeruchim2006communication}. Fitting $\nnfun$ to data
  generated by that model could, in principle, yield an estimator that effectively
  exploits the path loss and shadowing structure.
  \myitem\cmt{approach1}%
\begin{myitemize}%
  \myitem\cmt{training phase}%
    \begin{myitemize}%
      \myitem\cmt{Data generation}The idea is therefore to generate
      $T$ maps $\{ \receivedpsd_{t}(\bm x, f)\}_{t=1}^T$ together with
      $T$ sampling sets $\{\Omega_t\}_{t=1}^T$. Afterwards, $\{\tbm
      \truepsd_t\}_{t=1}^T$ and $\{\augmeaspsdmat_t\}_{t=1}^T$ can be
      formed as described earlier.  \myitem\cmt{denoising}It is
      possible to add artificially generated noise to the synthetic
      measurements in $\augmeaspsdmat_t$ to model the effect of
      measurement error. This would train the network to counteract
      the impact of such error, along the lines of  denoising autoencoders~\cite[Ch. 14]{goodfellow2016deep}.  \myitem\cmt{obj. func.}The advantage of this
      approach is that one has access to the ground truth, i.e., one
      can use the true maps $\bm \truepsd_{t}$ as
      \emph{targets}. Specifically, the neural network can be trained
      on the data $\{( \augmeaspsdmat_t, \bm
      \truepsd_{t})\}_{t=1}^{T}$ by solving
\begin{align}
\label{eq:objfunction1}
\textstyle\text{min.}_{~\bm w}\quad\frac{1}{T N_f}\sum_{t=1}^T
\sum_{f \in
  \mathcal{F}}  \left \Vert \bm  \truepsd_{t}(f)-
p_{\bm w}( \augmeaspsdmat_t(f) )\right \Vert_F^2.
\end{align}
    \end{myitemize}%
\myitem\cmt{test phase}If the model or simulator is sufficiently close
to the reality,  completing a real-world  map $\augmeaspsdmat(f)$  as
$\nnfun(\augmeaspsdmat(f))$ should produce an accurate estimate.
\end{myitemize}%
\myitem\cmt{challenge}
\end{myitemize}%

\subsubsection{Real Training Data}
\begin{myitemize}%
  \myitem\cmt{approach2}In practice, real maps may be available for
  training. However, in most cases, it will not be possible to collect
  measurements at all grid points before the map changes. Besides, it
  is not possible to obtain the entries of $\bm \truepsd$ but only
  measurements of it. This means that a real training  set is of
  the form $\{ \cbm \truepsd_t ,~t=1,\ldots,T\}$.
  \begin{myitemize}%
    
    \myitem\cmt{train}
    \begin{myitemize}%
      \myitem\cmt{Direct usage}For training, one can plug this data
      directly into \eqref{eq:fsep}. However, $\nnfun$ may then focus
      on learning just the values $\{ [\tbm \truepsd_t(f)]_{i,j},~
      (i,j)\in \Omega_t \}$, as would happen e.g. when $\nnfun$ is the
      identity mapping.  \myitem\cmt{Sample splitting}To counteract
      this trend, one can use one part of the measurements as the input and
      another part as the output (target).
      \begin{myitemize}%
        \myitem\cmt{index sets}For each $t$, construct the $Q_t$ pairs
        of (not necessarily disjoint)
        subsets $\Omega_{t,q}^{(I)},\Omega_{t,q}^{(O)}\subset
        \Omega_t$, $q=1,\ldots,Q_t$, e.g by drawing a given number of elements of
        $\Omega_t$ uniformly at random without replacement.
        \myitem\cmt{subsample}Using these subsets, subsample $\tbm
        \truepsd_t (f)$ to yield $\tbm \truepsd_{t,q}^{(I)}(f) \define
        \mathcal{P}_{\Omega^{(I)}_{t,q}}(\tbm \truepsd_{t}(f))$
        and $\tbm \truepsd_{t,q}^{(O)}(f) \define
        \mathcal{P}_{\Omega^{(O)}_{t,q}}(\tbm \truepsd_{t}(f))$. 
%
%
        \myitem\cmt{obj. func.}With these $TN_f \sum_tQ_t$ training
        examples, one can think of training as
        \begin{align}\label{eq:objfunction2}
          \begin{split}
          \minimize_{\bm w} & \textstyle\quad\frac{1}{TN_f \sum_tQ_t}\sum_{f \in
            \mathcal{F}}\sum_{t=1}^T \sum_{q=1}^{Q_t} \\&            
          \left\|
          \mathcal{P}_{\Omega_{t,q}^{(O)}}\left(
         \tilde {\bm  \truepsd}_{t,q}^{(O)}(f) - 
           p_{\bm w} \left(\augmeaspsdmat_{t,q}^{(I)}(f)\right)
          \right)
          \right\|_F^2,
          \end{split}
        \end{align}
      \end{myitemize}%
      where $\augmeaspsdmat_{t,q}^{(I)}(f)
$ has  $\tilde {\bm
  \truepsd}_{t,q}^{(I)}(f)$ and  $\bm \mask_{\Omega_{t,q}^{(I)}}$ as
frontal slabs.

    \end{myitemize}%
    \myitem\cmt{Limitations}
  \end{myitemize}%
\end{myitemize}%

\subsubsection{Hybrid Training}
In practice, one expects to have real data, but only in a limited
amount. It makes sense to apply the notion of \emph{transfer learning}~\cite[Ch. 15]{goodfellow2016deep} as follows: first, learn an initial parameter vector $\hbm
w$ by solving~\eqref{eq:objfunction1} with synthetic data. Second,
solve~\eqref{eq:objfunction2} with real data, but using $\hbm w$ as
initialization for the optimization algorithm. The impact of choosing
this initialization is that the result of
solving~\eqref{eq:objfunction2} in the second step will be closer to a
``better'' local optimum than if a worse initialization were adopted.

\subsection{Deep Completion Autoencoders}
\label{sec:convautoenc}
\cmt{Overview}This section proposes a deep neural network architecture
based on~\emph{convolutional autoencoders}~\cite{ribeiro2018study}.

\cmt{autoencoder}
  \begin{myitemize}%
    \myitem\cmt{conventional AE}A (conventional) autoencoder~\cite[Ch. 12]{goodfellow2016deep} is a neural
    network $\nnfun $ composed of two parts, an encoder $\encfun$ and a
    decoder $\decfun$, which satisfy
    $\nnfun(\auxinmat)=\decfun(\encfun(\auxinmat))~\forall\auxinmat$.
    The output of the encoder $\code\define\encfun(\auxinmat)\in
    \rfield^{\latentnum}$ is referred to as the \emph{code} or vector
    of \emph{latent variables} and is of a typically much lower
    dimension than the input $\auxinmat$. An autoencoder is trained so
    that $\decfun(\encfun(\auxinmat))\approx \auxinmat ~\forall
    \auxinmat$, which forces the encoder to compress the information
    in $\auxinmat$ into the $\latentnum$ variables in
    $\code$. The selection of $\latentnum$ will be addressed later.

    \myitem\cmt{Completion AE}
  \begin{myitemize}%
    \myitem\cmt{description}A \emph{completion} autoencoder adheres to
    the same principles as conventional autoencoders except for the
    fact that the encoder must determine the latent variables from a
    subset of the entries of the input. If a  mask is used,
    the transfer function must satisfy
    $
\auxinmat \approx \decfun(\encfun(\mathcal{P}_{\Omega}(\auxinmat),\bm
\mask_\Omega))~\forall\auxinmat
$
and for a sampling set $\Omega$ that preserves sufficient
information for reconstruction. If $\Omega$ does not satisfy this
requirement, then reconstructing $\auxinmat$ is impossible regardless
of the  technique used.  \myitem\cmt{this application}In the
application at hand and with the notation introduced in previous
sections, the above expression becomes
$\measpsdmat \approx \decfun(\encfun( \augmeaspsdmat ))$.

  \end{myitemize}%

  \end{myitemize}%

\begin{figure}[t!]
\centering
\includegraphics[width=1\columnwidth]{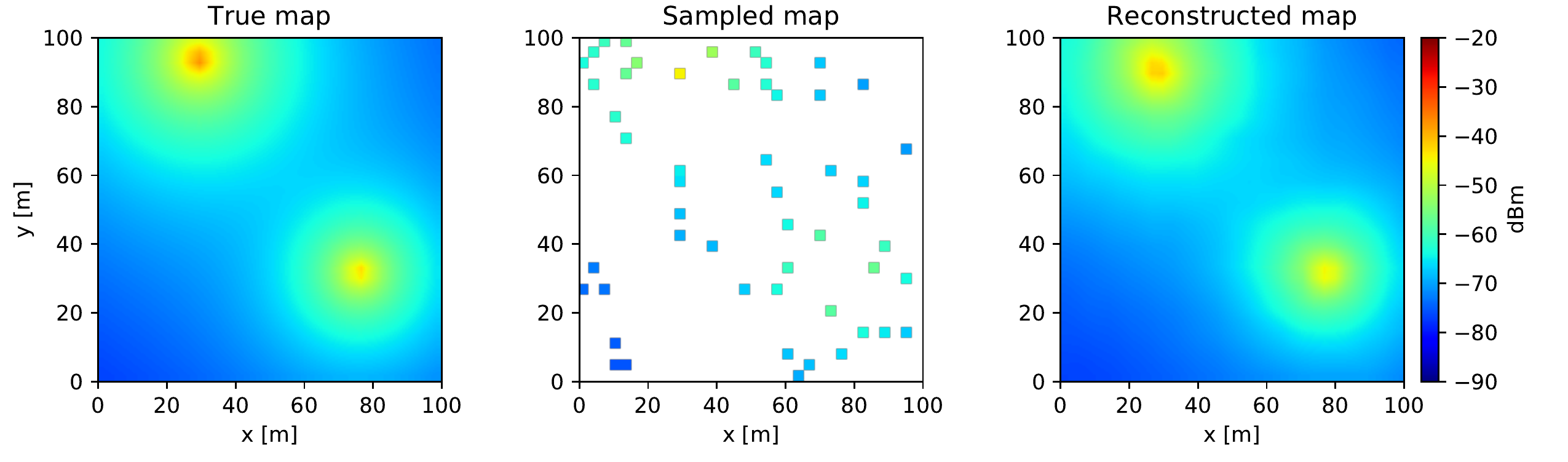}
\caption{Estimation with $N_\lambda=4$ latent
  variables: (left) true map, (middle) sampled map portraying 
   grid points $ \{\gridpoint_{i,j} \}$ with $\vert
  \mathcal{A}_{i,j}\vert > 0 $, and (right) estimated map. }
\label{f:motivatingex}
\end{figure}

\begin{figure}[t!]
\centering
\includegraphics[width=1\columnwidth, height=30\baselineskip]{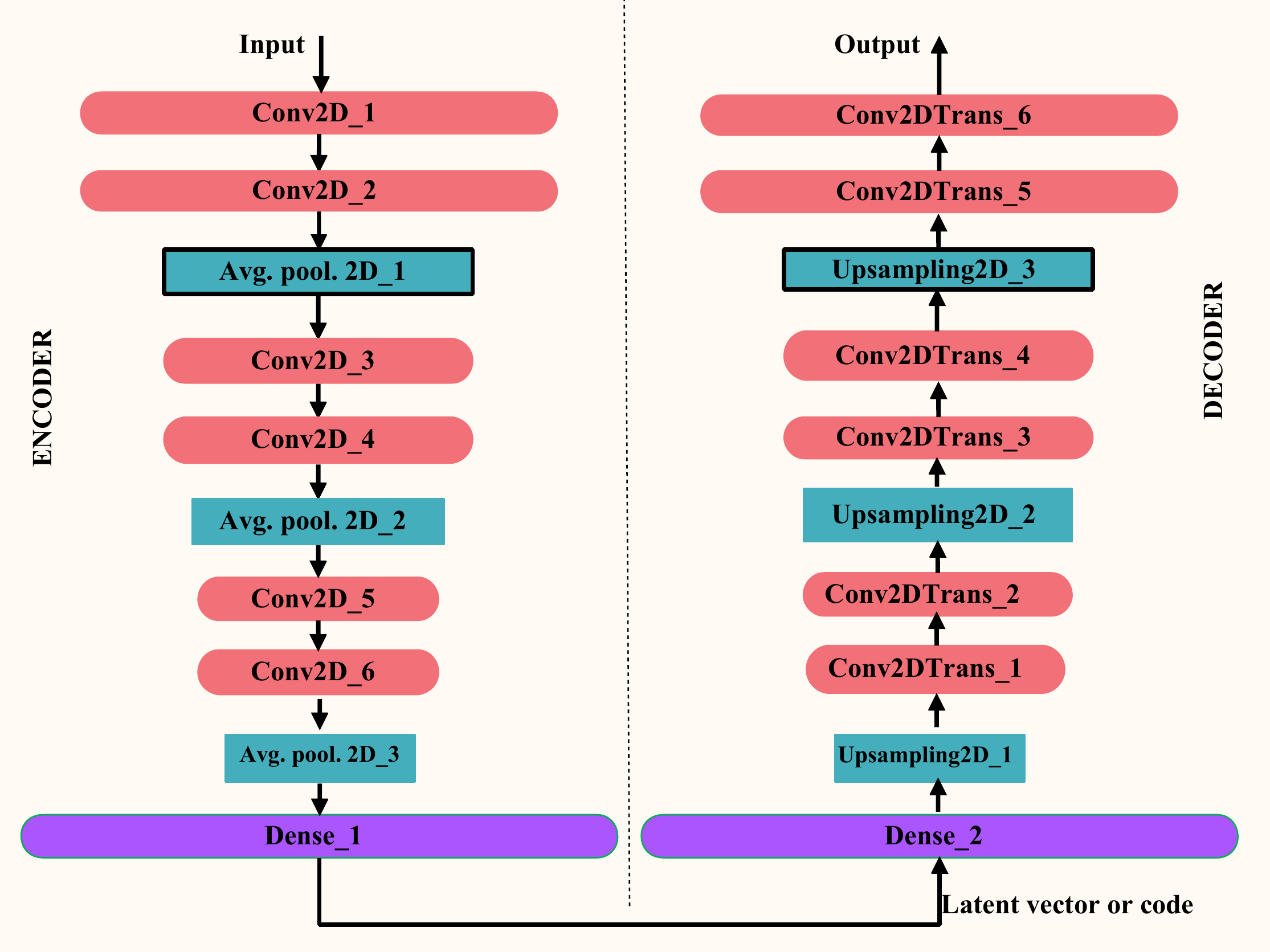}
\caption{Autoencoder architecture.} 
\label{f:modelarch}
\end{figure}

\cmt{latent variables}
\begin{myitemize}%
  \myitem\cmt{requirement for autoencoders}As indicated earlier,
  autoencoders are useful only when most of the information in the
  input can be condensed in $\latentnum$ variables, i.e., when the
  possible inputs lie close to a manifold of dimension $\latentnum$.
  \myitem\cmt{example}To see that this is the case in spectrum
  cartography, an illustrating toy example is presented next.
    \begin{myitemize}%
    \myitem\cmt{description}Suppose that there are two sources, each
    one with a fixed (yet possible different) power, that can be
    placed at arbitrary positions in $\mathcal{X}$ and suppose that
    propagation occurs in free space. All possible spectrum maps in
    this setup are defined by $\latentnum=4$ quantities, which
    correspond to the x and y coordinates of the two sources.
    \myitem\cmt{figure}Fig.~\ref{f:motivatingex} illustrates this
    effect, where the left panel of Fig.~\ref{f:motivatingex} depicts
    a true map $\bm \truepsd$ and the right panel shows its estimate
    using a completion autoencoder with $\latentnum=4$. The quality of
    the estimate clearly supports the aforementioned manifold
    hypothesis.  Details about the network and simulation setup are
    provided in Sec.~\ref{sec:numtest}.  \myitem\cmt{shadowing}In a
    real-world scenario, there may be more than two sources, their
    transmit power may not always be the same, and there are shadowing
    effects, which means that $\latentnum\geq 4$  will be
    required.
    \end{myitemize}%
\end{myitemize}%
\cmt{Architecture}The rest of this section will describe the main aspects of the architecture developed in this work
and summarized in Fig.~\ref{f:modelarch}.

\begin{table}
\normalsize
\renewcommand{\arraystretch}{1.25} 
\caption{Parameters of the proposed network.}
\label{table:autoencoder_par}
\centering 
\begin{tabular}{|p{2.7cm}|p{4.5cm}|}
\hline
\multirow{1}{8em}{ Layers}& Parameters \\ 
\hline
\multirow{1}{8em}{Conv2D/ Conv2DTranspose} & Kernel size = $3\times3$, stride = 1,
activation = PLReLU,   64 filters  \\ 
\hline
\multirow{1}{8em}{AveragePooling2D} & Pool size = 2, stride = 2  \\ 
\hline
\multirow{1}{8em}{Upsampling2D} & Up-sampling factor = $2$, bilinear interpolation \\  
\hline
\multirow{1}{8em}{Dense} & 64 neurons (encoder), 1024 neurons (decoder) \\  
\hline
\end{tabular}
\end{table}

  \begin{figure*}[t!]
    \centering\includegraphics[width=\textwidth]{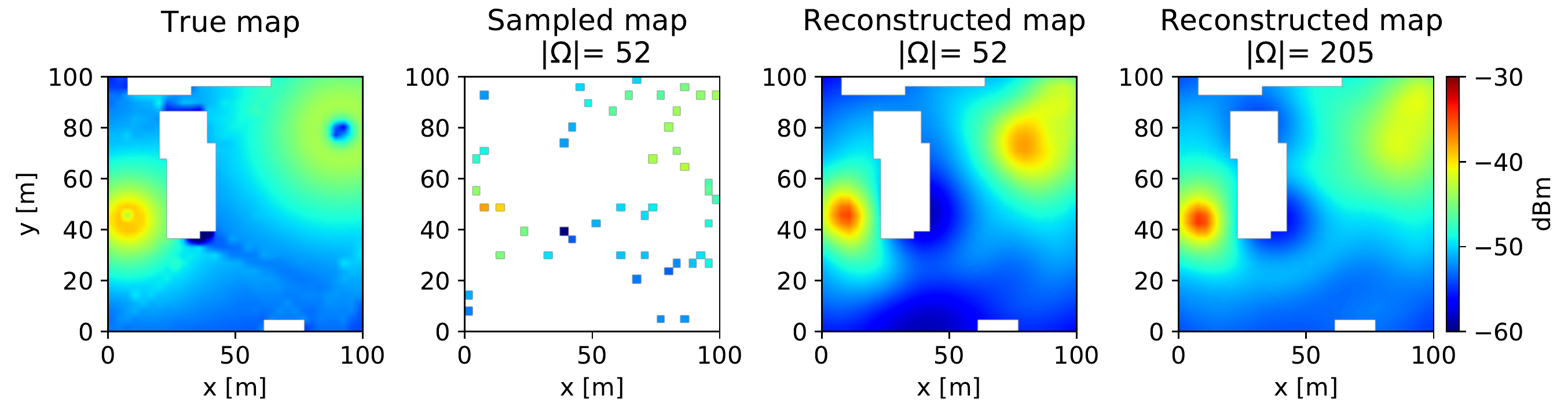}
    \caption{Power map estimate with the proposed neural network.
      (left): true map, (center left): sampled map portraying the locations
      of the grid points $ \{\gridpoint_{i,j} \}$ where $\vert
      \mathcal{A}_{i,j}\vert > 0 $; (center right) and (right):
      estimated maps. White areas represent buildings.  }
    \label{fig:reconstructionsample}
  \end{figure*}

\begin{myitemize}%
  \myitem\cmt{encoder}The encoder mainly comprises convolutional and
  pooling layers. 
  \begin{myitemize}%
    \myitem \cmt{convolutional layers}%
    \begin{myitemize}%
      \myitem\cmt{motivation}%
      The motivation for convolutional
      layers is three-fold:
      \begin{myitemize}%
        \myitem\cmt{motivation}
\cmt{reduce no. params. }(i) relative to fully connected
          layers, they severely reduce the number of parameters to train and, consequently,
          the amount of data required.
\cmt{shift-invariance \ra convolutional NN} Despite  this
          drastic reduction, (ii) convolutional layers are still capable of
          exploiting the spatial structure of maps and (iii) they result in
          shift-invariant transfer functions, a desirable property in the
          application at hand since moving the sources in a certain direction
          must be corresponded by  the same movement in the estimated
          map. 
      \end{myitemize}%
      \myitem\cmt{description}These layers compute
      \begin{align*}
        [\auxlayerout]_{i, j,\outchind} =\sum_{\inchind=1}^\inchnum \sum_{u=-k}^{k}\sum_{v=-k }^{k} [\bm F_\outchind]_{u, v,\inchind}[\auxlayerin]_{ i -u, j -v,\inchind},
      \end{align*}
      where
    \begin{myitemize}%
      \myitem $\auxlayerout$ is the output tensor,
      \myitem $\auxlayerin$ is the input tensor,
      \myitem and $\bm F_\outchind$ is the $\outchind$-th filter (or kernel),
      which is of size $2k+1 \times 2k+1$. 
    \end{myitemize}%
    Layer indices were omitted in order not to
    overload notation. 
      \myitem\cmt{activation \ra PReLUs}The activation functions used
      here are parametric
      \emph{leaky rectified linear units} (PLReLUs)~\cite{he2015delving} whose leaky parameter is also trained. 

    \end{myitemize}%
        \myitem\cmt{pooling}
    \begin{myitemize}%
\myitem\cmt{avgpool}On the other hand,~\emph{average pooling}
layers  down-sample the
outputs of convolutional layers, thereby condensing the information
gradually in fewer features. Additionally, pooling features
are approximately shift invariant as well~\cite[Ch. 9]{goodfellow2016deep}.
    \end{myitemize}
    
\myitem\cmt{dense}The last layer of the encoder is
\emph{fully-connected}. Since the previous layers were constrained to
be convolutional or pooling layers, a final fully-connected layer is
included in the encoder so that the latent variables can capture
arbitrary relations among the shift invariant features obtained by the
output of the second-to-last layer.
  \end{myitemize}%

  \myitem\cmt{decoder}As usual in autoencoders, the decoder follows a
  ``reverse'' architecture relative to the encoder. 
  \begin{myitemize}%
     \myitem\cmt{conv2DTr}Wherever the encoder has a
    convolutional layer, the decoder has a corresponding~\emph{
      convolution transpose} layer~\cite{dumoulin2016guide}, sometimes called ``deconvolutional'' layer.
    \myitem\cmt{upsampling}Likewise, the pooling layers of the
    encoder are matched with \emph{up-sampling} layers, which use  bilinear interpolation in the
    architectures that we investigated. 
    \myitem\cmt{dense}Finally, the fully connected  layer of the encoder
    is paired with a fully connected layer in the decoder. 
  \end{myitemize}%
\end{myitemize}%
\cmt{for layers \ra table}The overall network architecture is
summarized in Fig.~\ref{f:modelarch} and Table \ref{table:autoencoder_par}.

\begin{figure}[t!]
    \centering
    \includegraphics[width=1\columnwidth,height=10cm]{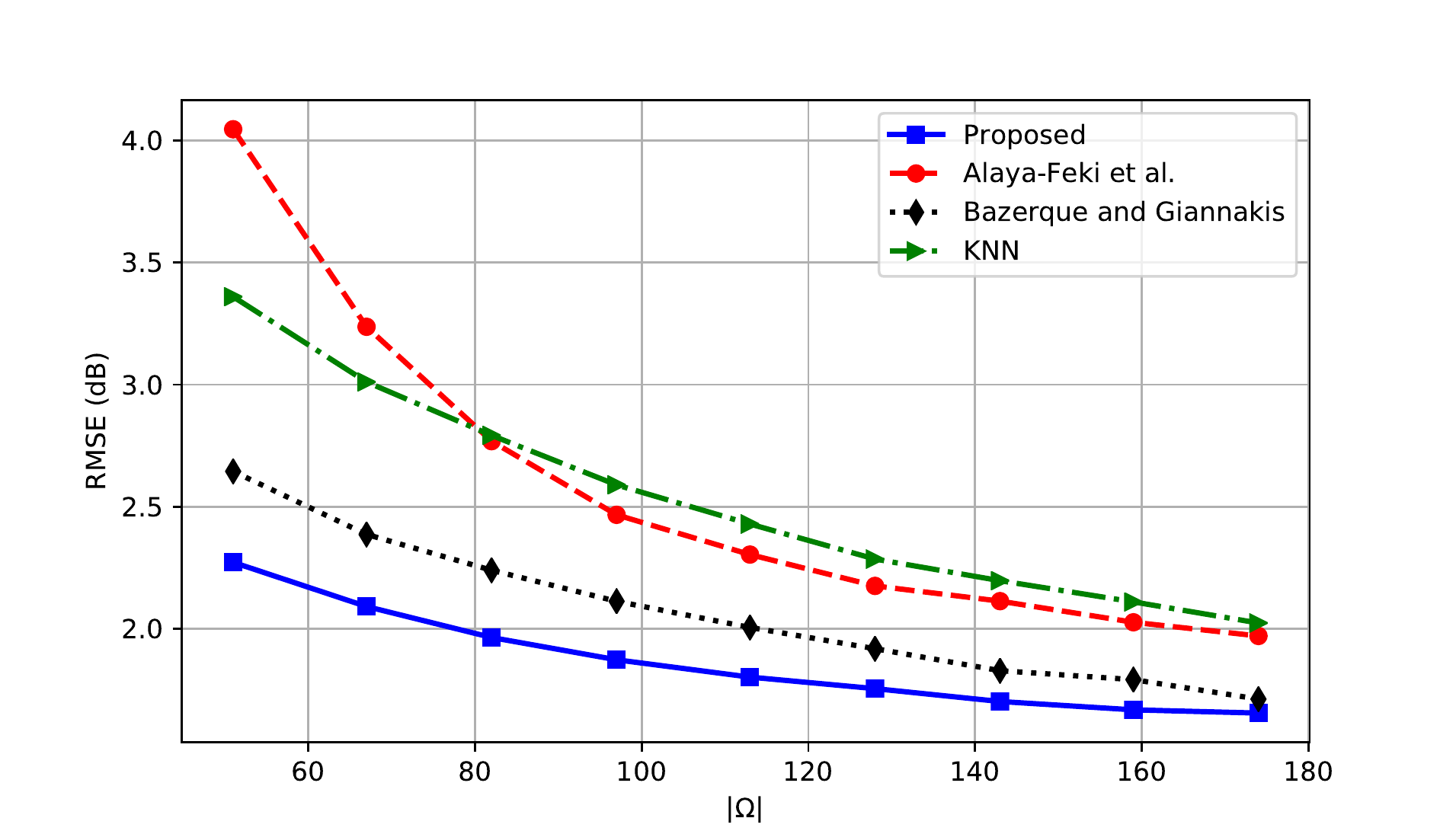}
    \caption{Comparison with state-of-the-art alternatives. Even
      though the  parameters of the competing algorithms were
      tuned for this specific experiment, the proposed network
      offers a markedly better performance. 
    }
    \label{f:comparison}
  \end{figure}
\begin{figure}
    \centering
    \includegraphics[width=0.8\columnwidth,height=9cm]{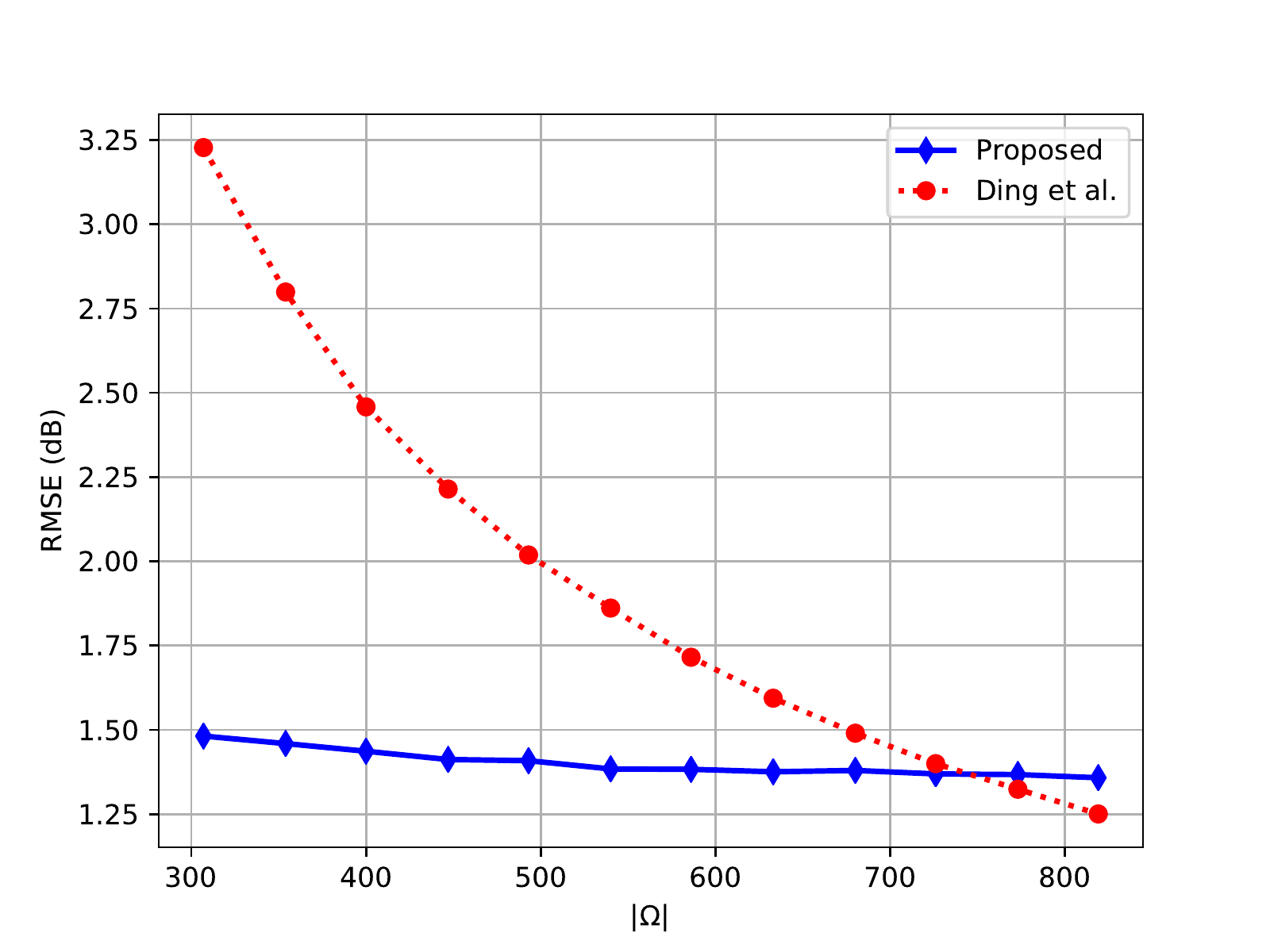}
    \caption{Performance comparison of the proposed scheme with that of the matrix completion algorithm in~\cite{ding2015devicetodevice}. The number of grid points in $\mathcal{X}$, $N_yN_x=1024$. 
    }
    \label{f:comparison_matrix}
  \end{figure}

\section{Numerical Experiments}
\label{sec:numtest}
\cmt{overview}This section validates the proposed framework and
network architecture through numerical experiments.
\cmt{data
  generation}
\begin{myitemize}%
  \myitem\cmt{freq}Thus, $\mathcal{F}$ is set to the singleton
  $\mathcal{F}=\{900 \text{ MHz}\}$.  \myitem\cmt{spatial
    region}$\mathcal{X}$ is a square area of side $100$ m, discretized
  into a grid with $N_y=N_x=32$.  \myitem\cmt{sources}The two
  considered transmitters have height 1.5 m and transmit power $11$
  and $7$ dBm over a bandwidth of 5 MHz.

  \myitem\cmt{Propagation
    channel}Two classes of maps are generated.
  \begin{myitemize}%
    \myitem\cmt{gudmundson}First, $T=4\cdot 10^5$ maps are obtained
  \begin{myitemize}%
    \myitem\cmt{source locations}where the two transmitters are placed uniformly at
    random 
    \myitem\cmt{channel}and where propagation adheres to the Gudmundson
     model~\cite{gudmundson1991correlation} with
    \begin{myitemize}%
      \myitem\cmt{pathloss exp}pathloss exponent 3,
      \myitem\cmt{gain unit distance}gain at unit distance $-30$ dB,
      \myitem\cmt{correlation}and shadowing correlation
      $\expected{H_\sourceind(\bm x_1,f)H_\sourceind(\bm
        x_2,f)}=\sigma_\text{sh}^20.95^{\vert\vert  \bm x_1 -  \bm x_2
        \vert\vert}$ with $\sigma_\text{sh}^2= 10$ dB$^2$.      
    \end{myitemize}
    \myitem\cmt{sensor locations}Sensors are distributed uniformly at
    random without replacement across the grid points. 
  \end{myitemize}
    \myitem\cmt{remcom}A separate set of maps is
    generated using Remcom's Wireless InSite software
    \begin{myitemize}%
  \myitem\cmt{channel}in an urban scenario.  \myitem\cmt{sensor
    locations}Sensors are distributed uniformly at random without
  replacement across the grid points that lie on the streets.
      \end{myitemize}
  \end{myitemize}%
  \myitem\cmt{noise psd}To better observe the impact of  propagation phenomena,
  $\upsilon(\bm x, f)$ is set to 0.  \myitem\cmt{measurements}Each
  measurement $ \tilde \receivedpsd(\gridpoint_{i,j}, f)$ is obtained
  by adding zero-mean Gaussian noise with standard deviation 1 dB to
  $\receivedpsd(\gridpoint_{i,j}, f), ~(i,j)\in\Omega$.
    
  \end{myitemize}

\cmt{Algorithms}
\begin{myitemize}%
  \myitem\cmt{Proposed}The network proposed in
  Sec.~\ref{sec:convautoenc} is implemented in TensorFlow and trained
  using the ADAM solver with learning rate $10^{-4}$. In this work, one training approach is analyzed, in this case
  \eqref{eq:objfunction1} with $\{( \augmeaspsdmat_t, \bm
  \truepsd_{t})\}_{t=1}^{T}$ the Gudmundson data set.
  \myitem\cmt{Competing}The algorithm is compared against the
  state-of-the-art competitors described next, whose parameters were
  tuned to approximately optimize their performance in the second
  experiment.
  \begin{myitemize}%
    \myitem\cmt{alaya-feki}(i) The kriging algorithm
    in~\cite{alayafeki2008cartography} with regularization parameter
    $10^{-5}$ and Gaussian radial basis functions with parameter
    $\sigma_{K} \define 3\sqrt{{\Delta_yN_y\Delta_xN_x}/{\vert
        \Omega\vert}}$, which is approximately 3 times the mean
    distance between two points at which measurements have been
    collected.  \myitem\cmt{multikernel}(ii) The multikernel algorithm
    in~\cite{bazerque2013basispursuit} with 20 Laplacian kernels with
    parameter uniformly spaced between $[0.1\sigma_{K},\sigma_{K}]$
    and regularization parameter $10^{-4}$.  \myitem\cmt{matrix
      completion}(iii) The matrix completion via nuclear norm minimization in~\cite{ding2015devicetodevice} with regularization parameter
    $10^{-5}$. \myitem\cmt{KNN}As a benchmark, (iv) the $K$-nearest
    neighbors algorithm  with $K=5$ is also shown.
  \end{myitemize}
\end{myitemize}
  
\cmt{Experiments}
\begin{myitemize}%
  \myitem\cmt{Experiment 1}The first experiment shows an estimated
  map using the proposed algorithm. The first panel of Fig.~\ref{fig:reconstructionsample} depicts the true
  map, which was generated using the Remcom data set. The second panel
  shows $\tbm \truepsd$ whereas the third and fourth show map estimates
  using different numbers of measurements. Observe that with just
  $|\Omega|=52$ measurements, the estimate is already of a high
  quality. Note that details due to diffraction or the directivity of
  the antennas are not reconstructed because the Gudmundson model used
  to train the network does not capture them and therefore the network
  did not learn these features. This illustrates the need for training
  over data sets that model the reality as close as possible. 

  \myitem\cmt{Experiment 2}The second experiment compares
  \begin{myitemize}%
  \myitem\cmt{metric}the root mean square error $ \text{RMSE}=
  \sqrt{{\mathbb{E}\{ \vert \vert \bm \truepsd-\hat{\bm \truepsd}\vert
      \vert_F ^2 \}}/{(N_y N_x)}}, $ of the aforementioned algorithms,
  where ${\bm \truepsd}$ is the true map, drawn at random from the
  Gudmundson data set, $\hat{\bm \truepsd}$ is the estimated map, and
  $\mathbb{E}\lbrace \cdot \rbrace $ denotes expectation over maps,
  noise, and sensor locations.  \myitem\cmt{Figure}From
  Fig.~\ref{f:comparison}, the proposed scheme performs approximately
  a 20 \% better than the next competing alternative. The parameters
  of the competing algorithms were tuned for this specific experiment,
  so their performance as in Fig.~\ref{f:comparison} is optimistic.
  In practice one must expect a greater performance gap.
 
  \end{myitemize}
  \myitem\cmt{Experiment 3}Due to poor performance of the matrix completion algorithm in~\cite{ding2015devicetodevice} in the adopted range of $|\Omega|$ in the second experiment, the third experiment compares
  \begin{myitemize}%
  \myitem\cmt{metric}its RMSE with that of the proposed algorithm in another range of $|\Omega|$ where there exists a higher number of measurements. \myitem\cmt{Figure}From Fig.~\ref{f:comparison_matrix}, the proposed method clearly outperforms the scheme in~\cite{ding2015devicetodevice} except when the number of measurements is very large, close to the total number of grid points. 
   \end{myitemize}
  \end{myitemize}

\section{Conclusions}
\label{sec:conclusion}
Learning propagation features from data yields spectrum cartography
algorithms that require fewer measurements to attain a target
performance. Deep neural networks can bring this idea into practice
and offer a performance that beats the state-of-the-art. Future work
will design more sophisticated network architectures relying on larger
data sets. 
\begin{myitemize}%
  \myitem\cmt{Summary}
  \myitem\cmt{Future work}  
\end{myitemize}%

\if\editmode1 
\onecolumn
\printbibliography
\else
\bibliography{\bibfilenames}
\fi
\end{document}